# CUMULATIVE OXYGEN ABUNDANCES OF SPIRAL GALAXIES

Stuart Dack and Marshall L. McCall
Department of Physics and Astronomy, York University, 4700 Keele Street, Toronto, ON M3J 1P3, Canada; studack@alumni.yorku.ca, mccall@yorku.ca

## ABSTRACT

Studying the global evolution of spiral galaxies requires determining their overall chemical compositions. However, since spirals tend to possess gradients in their chemical compositions, determining their overall chemical abundances poses a challenge. In this study, the framework for a newly proposed method for determining the overall oxygen abundance of a disk is established. By separately integrating the absolute amounts of hydrogen and oxygen out to large radii, the cumulative oxygen abundance is shown to approach an asymptotic value. In this manner, a reliable account of the overall chemical state of a disk is revealed.

## 1. INTRODUCTION

In order to fully understand the history of the universe, it is essential to understand the history of the individual galaxies within it. To understand the history of galaxies, it is necessary to learn about the types and amounts of elements of which they consist. Many factors may contribute to the evolution of galaxies, including initial conditions (Guesten & Mezger 1982), mass accretion (Chiosi 1980; Lacey & Fall 1985; Tsujimoto et al. 1995), star formation (Wyse & Silk 1989; Phillipps & Edmunds 1991), interactions of galaxies (Barnes & Hernquist 1991; Stewart et al. 2008), and the outflow of gas (Dahlem et al. 1998; Pettini et al. 2000). These factors may affect evolution locally or globally.

Determining the metallicity of a galaxy grants insight into its evolution because the elemental composition is directly related to star formation. The cumulative effect of all star formation during the history of the galaxy is seen in the total enrichment of heavy elements. Within each galaxy, local variations of metallicity may occur.

A readily observable gauge of metallicity in star-forming regions is the oxygen abundance (Searle 1971; Shields 1974). Oxygen lines are among the most prominent within spectra of H II regions, and techniques for analyzing them are well developed (Edmunds & Pagel 1984; McCall et al. 1985; Vílchez & Pagel 1988). Oxygen is created almost entirely through stellar processes, and most of it is created inside short-lived massive stars, which end their lives as Type II supernovae. Since the lifetimes of these stars are negligible on a cosmic timescale, the creation of oxygen can be said to occur at a rate proportional to that of the star formation rate. Thus, modeling of the evolution of the oxygen content of galaxies is simplified.

A problem exists in quantifying the oxygen abundance of a spiral galaxy. In a galaxy possessing a gradient in its chemical composition, the oxygen abundance at any given radius cannot generally be said to represent the chemical state of the galaxy. A simple method that is sometimes adopted is to merely choose a particular radius to characterize the entire galaxy (Zaritsky et al. 1994). Different galaxies can then be compared via chemical compositions at this chosen radius. However, the radius chosen is necessarily arbitrary and the corresponding composition may not truly represent the actual composition of the galaxy as a whole. Inevitably, comparisons among galaxies are fraught with danger.

Sadavoy & McCall (2006) were the first to introduce the idea that the ratio of the global abundance of oxygen relative to hydrogen within a spiral galaxy might be determinable. This was done for two spirals, NGC 5457 and NGC 6946. They calculated the total number of oxygen atoms out to a certain distance from the center and divided it by the total number of hydrogen atoms within that distance. They observed that the ratio of the total number of oxygen atoms to the total number of hydrogen atoms approached a constant at large radii. The reason why the oxygen abundance approaches an asymptote is as follows. In the inner parts of a galaxy, $n(O)/n(H)$ is high. However, oxygen is spread over a small volume. The outer parts of a galaxy have less oxygen, but it is spread over a larger volume. As it turns out, the rate of decline of $n(O)/n(H)$ with radius wins out over the rate of increase in surface area of the disk, with the result that $n(O)/n(H)$ within some radius approaches an asymptote as the radius becomes large.

The first goal of the research presented in this study is to test whether the cumulative abundances for a large sample of spirals systematically approach asymptotes at large radii. The second goal is to analyze and understand the shape of the cumulative oxygen abundance (COA) curves, and thereby develop a way of robustly quantifying the asymptotic abundances.

This paper is organized in the following manner. In Section 2, methods for measuring oxygen abundances are examined, and the best indirect method for estimating oxygen abundances is established. In Section 3, the sample of galaxies for study is constructed. In Section 4, the methods of data reduction for oxygen abundance analyses are described. In Section 5, radial profiles of oxygen abundances for H II regions are presented for all galaxies in the study, and an analysis of these plots is performed. In Section 6, the oxygen abundance data are combined with measurements of hydrogen within each galaxy to compute amounts of oxygen and hydrogen as a function of radius. In Section 7, the COA curves are plotted and asymptotes are established. The main results are summarized in Section 8.

## 2. ANALYSIS OF NEBULAR SPECTRA

### 2.1. Direct and Indirect Methods for Calculating Oxygen Abundances

H II regions are often modeled as two distinct zones of ionization, one of low ionization where $O^+$ is the predominant form of oxygen, and one of high ionization where $O^{++}$ predominates.

An electron temperature is associated with each of these zones. In this paper, the temperatures for the O$^+$ and O$^{++}$ zones are referred to as T[O II] and T[O III], respectively. Models show that temperatures in the two zones of ionization are related to one another (Pagel et al. 1992; Izotov et al. 1997; Deharveng et al. 2000). The model presented by Garnett (1992) has been adopted here. It relates the two zones via the following equation:

$$\text{T[O II]} = 0.7\,\text{T[O III]} + 3000 \text{ K}. \quad (1)$$

The most effective method of finding abundances is to first measure the temperature in at least one of the zones. From there, an estimate of the abundances of O$^+$ and O$^{++}$ can be accomplished. This method is referred to as the "direct" method.

Models of statistical equilibrium can give a good estimate of the electron temperature in the O$^{++}$ zone from a comparison of the line flux at [O III] $\lambda 4363$ and one of the other doubly ionized oxygen ions, [O III] $\lambda 4959$ or [O III] $\lambda 5007$.

A challenge arises if the oxygen abundance of a region is larger than approximately 12+log(O/H) = 8.5 because [O III] $\lambda 4363$ can become too faint to be observed. Therefore, it is necessary to find another method for calculating oxygen abundances at high metallicities. Many techniques have been proposed, the most popular of which employs the combination of forbidden lines of oxygen in the $R_{23}$ index (Pagel et al. 1979)

$$R_{23} = \frac{([\text{O II}]\,\lambda 3727 + [\text{O III}]\,\lambda\lambda 4959, 5007)}{\text{H}\beta}, \quad (2)$$

where H$\beta$ refers to the hydrogen Balmer series line at 4861.32 Å. The forbidden line [O II] $\lambda 3727$ is actually a combination of the close doublet [O II] $\lambda\lambda 3726, 3729$.

A comparison of $R_{23}$ versus oxygen abundance shows that for oxygen abundances between approximately 7.95 and 8.20, no correlation exists (Pilyugin 2000, 2001). Therefore, when analyzing galaxies using a calibration of the $R_{23}$ index, it is necessary to restrict the sample of H II regions to those with oxygen abundances higher than 12+log(O/H) = 8.20. Typically, H II regions within large spiral galaxies do not have oxygen abundances that fall below 12+log(O/H) = 7.95. Methods for calculating oxygen abundances that make use of the $R_{23}$ index are referred to as "indirect" methods. Some such methods will be discussed in Section 3.

### 2.2. Density Determinations

In general, H II regions have relatively low electron densities. Typically, densities range from 10 cm$^{-3}$ to 250 cm$^{-3}$. Within this range, density variations barely affect statistical equilibrium. Over the range from 10 cm$^{-3}$ to 250 cm$^{-3}$, calculated oxygen abundances change by less than 0.01 dex. This variation is more than an order of magnitude smaller than typical errors of measurement. In this study, all densities are assumed to be 100 cm$^{-3}$.

### 3. CALCULATION OF OXYGEN ABUNDANCES

#### 3.1. Options for $R_{23}$ Calibration

Many different strong line calibration techniques have been proposed since the original $R_{23}$ index was introduced by Pagel et al. (1979). Three of the most widely used calibration methods, McGaugh (1991), Zaritsky et al. (1994), and Pilyugin (2001), were used in the study by Sadavoy & McCall (2006). The calibrations for calculating oxygen abundances are, respectively, as follows:

McGaugh (1991):

$$\begin{aligned} 12 + \log(\text{O/H}) = {} & 12 - 2.939 - 0.2x - 0.237x^2 \\ & - 0.305x^3 - 0.0283x^4 \\ & - y(0.0042 - 0.0221x - 0.102x^2 \\ & - 0.0817x^3 - 0.00717x^4), \end{aligned} \quad (3)$$

where $x = \log(R_{23})$ and $y = \log\frac{[\text{O III}]\,\lambda\lambda 4959, 5007}{[\text{O II}]\,\lambda 3727}$;

Zaritsky et al. (1994):

$$\begin{aligned} 12 + \log(\text{O/H}) = {} & 9.265 - 0.33x \\ & - 0.202x^2 - 0.207x^3 - 0.333x^4, \end{aligned} \quad (4)$$

where $x = \log(R_{23})$;

Pilyugin (2001):

$$12 + \log(\text{O/H}) = \frac{R_{23} + 54.2 + 59.45P + 7.31P^2}{6.07 + 6.71P + 0.37P^2 + 0.243R_{23}}, \quad (5)$$

where $P = \frac{[\text{O III}]\,\lambda\lambda 4959, 5007}{[\text{O II}]\,\lambda 3727 + [\text{O III}]\,\lambda\lambda 4959, 5007}$.

The McGaugh (1991) and Pilyugin (2001) calibrations have the advantage of including the ratio of [O II] $\lambda 3727$ to [O III] $\lambda\lambda 4959, 5007$ along with the $R_{23}$ index. Without incorporating this ratio, it is possible to misinterpret a spectral difference arising from a deviation in physical conditions as a consequence of a difference in chemical composition.

#### 3.2. Empirical Comparison of Direct and Indirect Abundances

Determining a proper calibration of oxygen abundances using indirect methods is done by comparing the results of the various calibrations to those obtained by the direct method. For H II regions whose oxygen abundance levels are below 12+log(O/H) = 8.20, this approach is not possible. At high metallicities (12+log(O/H) > 8.50), the [O III] $\lambda 4363$ auroral line becomes too faint to detect. Then, verification of indirect methods of calibration can be performed using planetary nebulae (discussed in Section 3.3).

In this study, entire galaxies were chosen to anchor calibrations as opposed to individual H II regions from a large variety of galaxies. This is because abundance gradients can be used to guide the connection of $R_{23}$ and direct abundances. The reduction of data was performed with SNAP (Spreadsheet Nebular Analysis Package), a program developed at York University in Toronto, Canada (Komljenovic et al. 1996).

Figure 1 shows abundances as a function of radius for two different galaxies, NGC 300 and NGC 5457, for which both the direct method and various indirect methods using $R_{23}$ have been applied. In these plots, the slopes of the gradients from the various calibration methods have been constrained to match the slope of the gradient of the direct method. In this way, it is possible to get an overall picture of the systematic difference in abundances from different calibrations. The difference in O/H between the various methods of calibration and the direct method are listed in Table 1. It is evident that, of the three methods of calculating oxygen abundances, the closest approximation is that of Pilyugin (2001).

In Table 1, it is clear that although the Pilyugin equation delivers abundances closest to the true values, there is still a small difference. Nonetheless, for this paper the Pilyugin calibration was adopted in its original form.

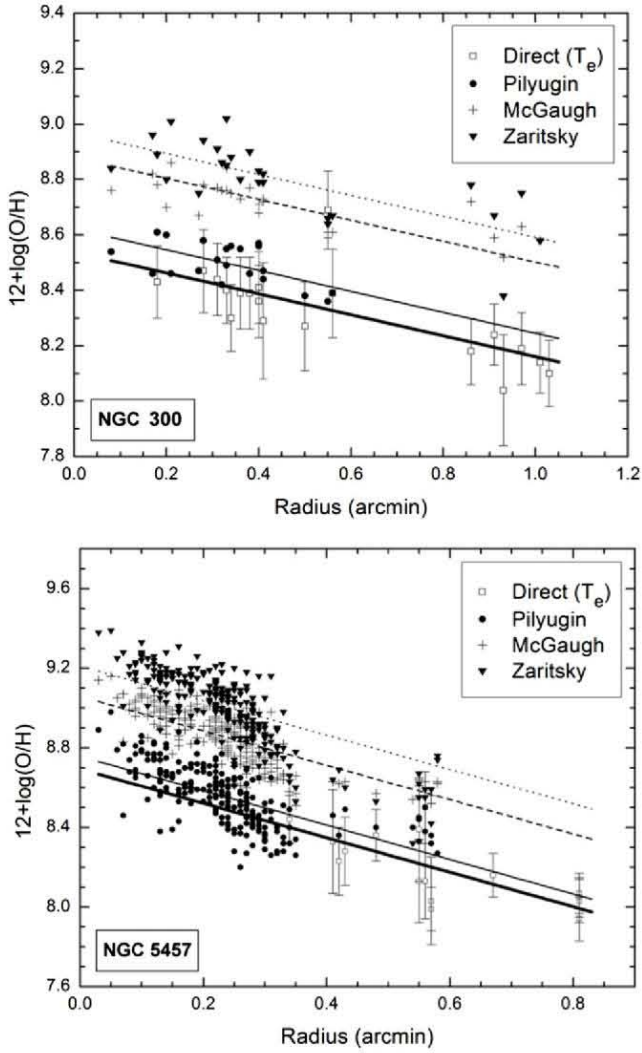

**Figure 1.** Oxygen abundance vs. deprojected radius for NGC 300 (top) and NGC 5457 (bottom). In each graph, the thick solid line represents the best linear fit to abundances derived from the direct method. The thin solid line, the dashed line, and the dotted line represent the best linear fits to the abundances from the calibrations of Pilyugin (2001), McGaugh (1991), and Zaritsky et al. (1994), respectively.

**Table 1**
Comparison of Various Calibration Methods for Determining Oxygen Abundances

| Calibration | | NGC 300 | NGC 5457 |
|---|---|---|---|
| | Number of H II regions with [O III] $\lambda 4363$ | 18 | 17 |
| Direct | Slope (dex arcmin$^{-1}$) | −0.38 | −0.86 |
| | Intercept (O/H at $r = 0$) | 8.54 | 8.69 |
| Pilyugin | Intercept (O/H at $r = 0$) | 8.62 | 8.76 |
| | Difference from direct | 0.08 | 0.07 |
| McGaugh | Intercept (O/H at $r = 0$) | 8.88 | 9.06 |
| | Difference from direct | 0.34 | 0.37 |
| Zaritsky | Intercept (O/H at $r = 0$) | 8.97 | 9.21 |
| | Difference from direct | 0.43 | 0.52 |

### 3.3. Verification of $R_{23}$ Calibration with Planetary Nebulae

Another possible approach to finding chemical abundances is by observing planetary nebulae. Planetary nebulae result from the death of long-lived stars, so they have oxygen abundances approximately equal to that of their progenitors (with little or no self-enrichment). Thus, depending on the star formation rate since the birth of the progenitors, planetary nebulae may undervalue the oxygen abundance in the interstellar medium (ISM; McCall & Richer 2003). On the other hand, H II regions are thought to have chemical compositions closest to that of the ISM (McCall & Richer 2003).

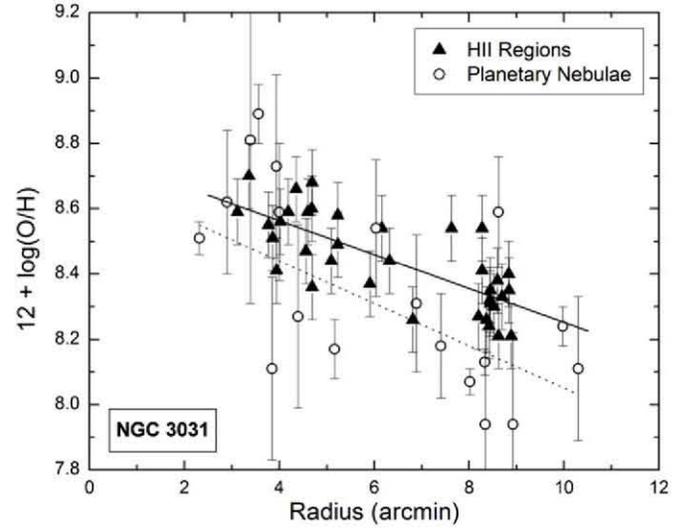

**Figure 2.** Oxygen abundance vs. deprojected radius for H II regions (closed triangles) and planetary nebulae (open circles) in NGC 3031. Abundances in H II regions were derived using the Pilyugin (2001) calibration. The solid and dotted lines represent, respectively, the best-fit trends for H II regions and planetary nebulae.

Although planetary nebulae may undervalue the oxygen abundance in the ISM (McCall & Richer 2003), with enough statistics, planetary nebulae may still be used to gauge oxygen abundances. Richer et al. (1997) showed that there is a relationship between planetary nebulae and H II region abundances at a given radius. Gradients of the oxygen abundance found for H II regions and planetary nebulae for the same galaxy have the same slope, but the trends can be offset from each other by a small amount, with H II regions displaying higher abundance values at all radii. The offset amounts to less than 0.15 dex, so it is possible to use planetary nebulae to check H II region determinations. This is particularly valuable where the oxygen abundance is high; H II regions are so cool that abundances cannot be determined without the aid of models, whereas central stars of planetary nebulae are so hot that auroral lines remain strong.

Figure 2 shows a comparison of the abundances of planetary nebulae and H II regions in NGC 3031. The solid line represents the best fit for the H II regions, while the dotted line is the trend for planetary nebulae. Note that the slopes of the best-fit lines (−0.052 and −0.065, respectively) are the same within the errors (0.008 and 0.015, respectively). With the slope for H II regions fixed at the value for planetary nebulae, the trends are separated by only 0.10 dex. This fact is important because it gives confidence to the high oxygen abundance end of the Pilyugin calibration, where H II regions are too cool to detect auroral lines, preventing the measurement of temperatures and the direct determination of abundances. The offset of H II region abundances from those for planetary nebulae is consistent with the abundance gap predicted for those metallicities by Richer et al. (1997).

Table 2
Physical Attributes of Galaxies in the Study

| Galaxy | Other Names | Type (1) | Dist. Mod. (2) | $R_O$ (′) (3) | $R_O$ (kpc) (4) | $i$ (°) (5) | P.A. (°) (6) |
|---|---|---|---|---|---|---|---|
| NGC 224 | M31, Andromeda | SA(s)b | 24.34 ± 0.09 | 102.09 | 23.5 | 78 | 38 |
| NGC 300 | Caldwell 70 | SA(s)d | 26.44 ± 0.08 | 11.19 | 86.1 | 46 | 107 |
| NGC 598 | M33, Triangulum | SA(s)cd | 24.62 ± 0.08 | 37.07 | 9.1 | 54 | 23 |
| NGC 628 | M74 | SA(s)c | 29.64 ± 0.07 | 5.36 | 46.2 | 9 | 48 |
| NGC 925 | | SAB(s)d | 29.78 ± 0.05 | 5.48 | 47.5 | 61 | 109 |
| NGC 1097 | | SB(s)b | 31.11 ± 0.20 | 4.77 | 43.2 | 31[a] | 141[a] |
| NGC 1232 | | SAB(rs)c | 31.34 ± 0.20 | 3.71 | 33.8 | 30[b] | 108[b] |
| NGC 1365 | Great Barred Spiral | SB(s)b | 31.25 ± 0.06 | 5.61 | 51.0 | 40 | 32 |
| NGC 2403 | Caldwell 7 | SAB(s)cd | 27.50 ± 0.05 | 11.45 | 91.6 | 60 | 125 |
| NGC 2903 | | SAB(rs)bc | 29.62 ± 0.19 | 6.29 | 54.2 | 61 | 23 |
| NGC 3031 | M81 | SA(s)ab | 27.75 ± 0.09 | 13.77 | 111.2 | 57 | 151 |
| NGC 3184 | | SAB(rs)cd | 30.22 ± 0.20 | 3.71 | 32.6 | 16[c] | 179[c] |
| NGC 3198 | | SB(rs)c | 30.64 ± 0.08 | 4.26 | 38.0 | 72 | 216 |
| NGC 4258 | M106 | SAB(s)bc | 29.29 ± 0.09 | 9.31 | 79.3 | 67 | 150 |
| NGC 4559 | | SAB(rs)cd | 29.03 ± 0.20 | 5.48 | 46.3 | 67[d] | 323[d] |
| NGC 5194 | M51, Whirlpool | SA(s)bc | 29.41 ± 0.11 | 5.61 | 48.0 | 20 | 166 |
| NGC 5236 | M83, Southern Pinwheel | SAB(s)c | 28.35 ± 0.10 | 6.59 | 54.3 | 25 | 66 |
| NGC 5457 | M101, Pinwheel | SAB(rs)cd | 29.24 ± 0.13 | 14.42 | 122.6 | 21 | 43 |
| NGC 6946 | Caldwell 12, Fireworks | SAB(rs)cd | 28.83 ± 0.22 | 8.30 | 69.6 | 33 | 66 |
| NGC 7793 | | SA(s)d | 27.78 ± 0.24 | 4.77 | 38.5 | 52 | 105 |

**Notes.** (1) Revised Hubble Type, from de Vaucouleurs et al. (1991). (2) Distance modulus and error, in mag, with the zero point set by the maser distance to NGC 4258, from M. L. McCall (2012, in preparation). (3) Isophotal radius, corrected to face-on and corrected for galactic extinction, in arcminutes, from de Vaucouleurs et al. (1991). (4) Same as (3), but in kiloparsecs, from de Vaucouleurs et al. (1991). (5) Inclination, from M. L. McCall (2012, in preparation). (6) Position angle measured eastward from north, from M. L. McCall (2012, in preparation).
[a] Ondrechen et al. (1989).
[b] van Zee & Bryant (1999).
[c] Begeman (1989).
[d] Barbieri et al. (2005).

## 4. SELECTION OF GALAXIES

### 4.1. Selection Criteria

Twenty galaxies were chosen for study. Galaxies at either extreme of the Hubble sequence were excluded. Elliptical galaxies have little gas and so do not contain H II regions. Also, their planetary nebulae are difficult to observe. Thus, their abundance gradients are not well known. Irregular galaxies, such as dwarfs, were excluded since their chemical compositions are homogeneous (Olszewski et al. 1991; Bica et al. 1998; Parisi et al. 2009; Cioni 2009) and they therefore do not generally possess a gradient in their oxygen abundance. The galaxies chosen in this sample range from 2.4 to 7.4 on the Revised Hubble System T scale (de Vaucouleurs et al. 1991).

The two galaxies studied by Sadavoy & McCall (2006) were included in order to compare and evaluate their findings. All analyses involving these two galaxies were re-done using self-consistent methods for data reduction and computations.

Distances were determined preferentially from stellar indicators, namely, Cepheid variables, planetary nebulae, the tip of the red giant branch, or surface brightness fluctuations. A pairwise analysis of 127 distances to 34 nearby galaxies was undertaken to define zero points for the four methods anchored to the maser distance to NGC 4258 (a distance modulus of 29.29: Herrnstein et al. 1999; Gibson 2000; Macri et al. 2006). The same galaxies were used to define the Tully–Fisher relation in $I$, which was employed for galaxies lacking observations of stellar constituents. Throughout, corrections for extinction and motion were accomplished with the aid of the York Extinction Solver (McCall 2004). The reddening law of Fitzpatrick (1999) was employed, set up to deliver a ratio of total-to-selective extinction of 3.07 for Vega (see McCall 2004). Further details concerning distance analyses are provided in Fingerhut et al. (2007) and M. L. McCall (2012, in preparation).

### 4.2. Published Information

For this study it was important to select galaxies for which there was published information on many H II regions located at a wide range of radii across each. Specifically, measurements of the emission lines of oxygen and hydrogen in H II regions and maps of 21.1 cm radiation from H I (which traces neutral hydrogen) were needed. In addition, galaxies with data relating to their molecular hydrogen content were favored. Such information comes in the form of carbon monoxide maps, since CO traces molecular hydrogen (Dickman 1978).

All data used in this analysis were collected from published works. The physical attributes of all galaxies in the study are listed in Table 2. Table 3 gives the sources from which the data for each galaxy were taken.

## 5. MEASUREMENTS OF OXYGEN ABUNDANCE GRADIENTS IN GALAXIES

In order to calculate oxygen abundances for H II regions within the sample galaxies, both the indirect method and, where possible, the direct method were used. For H II regions where [O III] $\lambda$4363 was available, only the direct method was applied so as not to double count the same region. It was possible to calculate oxygen abundances for NGC 300 solely using the direct method. For the galaxies NGC 5236 and NGC 5457, a combination of direct and indirect methods was used to calculate oxygen abundances. Abundances for the remaining galaxies in this study were obtained strictly using the Pilyugin (2001) $R_{23}$ method.

**Table 3**
Published Sources from Which Data Were Obtained

| Galaxy | O/H | H I | CO |
| --- | --- | --- | --- |
| NGC 224 | Blair et al. (1982) | Chemin et al. (2009) | Nieten et al. (2006) |
| NGC 300 | Bresolin et al. (2009a) | Westmeier (2011) | |
| NGC 598 | Bresolin (2011) | Newton (1980) | Corbelli (2003) |
| NGC 628 | McCall et al. (1985)<br>Ferguson et al. (1998)<br>van Zee et al. (1998)<br>Bresolin et al. (1999) | Wevers (1984) | Leroy et al. (2009) |
| NGC 925 | Dors & Copetti (2005) | Pisano et al. (1998) | Leroy et al. (2009) |
| NGC 1097 | Storchi-Bergmann et al. (1996) | Crosthwaite (2002) | Gerin et al. (1988) |
| NGC 1232 | van Zee & Bryant (1999)<br>Bresolin et al. (2005) | van Zee & Bryant (1999) | |
| NGC 1365 | Bresolin et al. (2005) | Holwerda et al. (2005) | |
| NGC 2403 | McCall et al. (1985)<br>van Zee et al. (1998)<br>Garnett et al. (1997)<br>Bresolin et al. (1999) | Leroy et al. (2008) | Thornley & Wilson (1995) |
| NGC 2903 | van Zee et al. (1998)<br>Bresolin et al. (2005) | Wevers (1984) | Young et al. (1995) |
| NGC 3031 | Garnett & Shields (1987)<br>Bresolin et al. (1999)<br>Stanghellini et al. (2010) | Holwerda et al. (2005) | Sage & Westpfahl (1991) |
| NGC 3184 | McCall et al. (1985)<br>Zaritsky et al. (1994)<br>van Zee et al. (1998) | Leroy et al. (2008) | Leroy et al. (2009) |
| NGC 3198 | Zaritsky et al. (1994) | Begeman (1989) | Leroy et al. (2009) |
| NGC 4258 | Bresolin (2011) | Wevers (1984) | Young et al. (1995) |
| NGC 4559 | Zaritsky et al. (1994) | Barbieri et al. (2005) | |
| NGC 5194 | Bresolin et al. (1999)<br>Bresolin et al. (2004) | Schuster et al. (2007) | Schuster et al. (2007) |
| NGC 5236 | Bresolin et al. (2005)<br>Bresolin et al. (2009b) | Rogstad et al. (1974) | Andersson (2002) |
| NGC 5457 | Kennicutt et al. (2003)<br>Bresolin (2007)<br>van Zee et al. (1998)<br>Cedrés & Cepa (2002) | Kenney et al. (1991) | Kenney et al. (1991) |
| NGC 6946 | McCall et al. (1985)<br>Ferguson et al. (1998) | Tacconi & Young (1986) | Leroy et al. (2009) |
| NGC 7793 | Webster & Smith (1983)<br>McCall et al. (1985) | Leroy et al. (2008) | |

The two galaxies sampled by Sadavoy & McCall (2006), NGC 5457 and NGC 6496, were included in this study. However, compared to that study, the sample of H II regions was enlarged and the analysis was improved. H II regions that lie below 12+log(O/H) = 8.20 were eliminated. Moreover, only the best choice for indirect oxygen abundance calibrations was used, namely, that of Pilyugin (2001).

Figure 3 shows plots of the oxygen abundance versus deprojected angular radius for all galaxies in this study. Sources of spectroscopy are listed in a legend within each graph. Since it is known that $n(O)/n(H)$ decays exponentially with radius (Searle 1971; Zaritsky 1992), plots of 12+log(O/H) versus radius can justifiably be fitted with a line. Errors for abundances derived from [O III] λ4363 were propagated from uncertainties in line fluxes using SNAP. Errors for abundances derived from the $R_{23}$ method are about 0.1 dex (Pilyugin 2001). In each plot, the thick solid line represents the oxygen abundance gradient, while the dashed lines represent the maximum and minimum gradients allowed by the data. These gradient limits were calculated by adding or subtracting the standard error of each slope to the slope itself, and then fitting the intercept.

## 6. OVERALL QUANTITIES OF HYDROGEN AND OXYGEN

### 6.1. Motivation

The gradients calculated in Section 5 enable the computation of COAs, namely, the total number of oxygen atoms divided by the total number of hydrogen atoms within any radius. In order

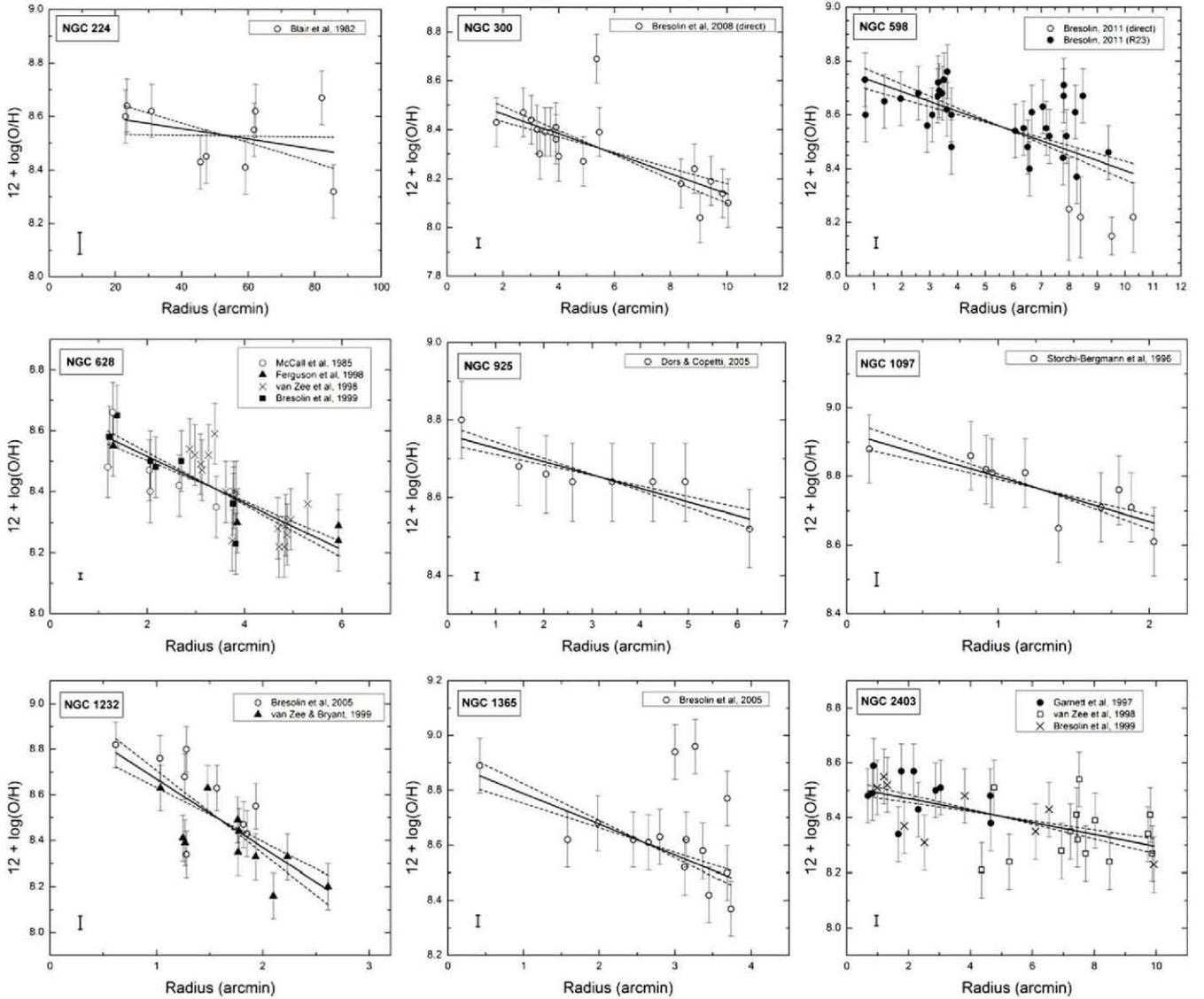

**Figure 3.** Oxygen abundance vs. deprojected radius for all galaxies in the sample. In each plot, the source from which the spectroscopic data were obtained is listed in the legend. Data points obtained using the direct method are indicated by the word "direct" next to the listed source. For sources from which points were obtained using the direct method as well as the $R_{23}$ method, points obtained using the $R_{23}$ method are indicated by "R23" next to the source. All other data points were obtained via the $R_{23}$ method. The least-squares best fit to the points is shown as a thick solid line. The dashed lines represent maximum and minimum fits calculated by adding or subtracting the standard error of each slope to the slope itself, and then fitting the intercept. The vertical error at the intersection is shown in the bottom-left corner.

to examine how cumulative abundances vary with radius, it is necessary to track as a function of radius:

1. the column density of hydrogen atoms;
2. the oxygen abundance.

Together, the total number of oxygen atoms within any radius is constrained. Oxygen abundances have already been determined for each galaxy in Section 5. Therefore, only the hydrogen is left to be found. In order to find the total hydrogen column density, it is necessary to calculate the column densities of all types of hydrogen:

$$N(\mathrm{H}) = N(\mathrm{H}^0) + 2N(\mathrm{H}_2) + N(\mathrm{H}^+). \qquad (6)$$

The column density of molecular hydrogen is multiplied by 2 to account for the two hydrogen atoms from which it is made. It can be assumed that $N(\mathrm{H}^+)$ is approximately zero, since it is restricted to H II regions or the hot ISM (which has a very small density).

### 6.2. Calculating the Quantities of Neutral and Molecular Hydrogen

The majority of atomic hydrogen within each galaxy is neutral, and its column density can be mapped via emission at 21.1 cm. The sources for H I data for all galaxies in the sample are listed in Table 3.

Unlike neutral hydrogen, molecular hydrogen, $H_2$, does not produce any easily observable emission lines. Fortunately, there exists a direct correlation between the column density of carbon monoxide, CO, and the column density of molecular hydrogen (Lebrun et al. 1983; Strong et al. 1988; Désert et al. 1988; Dame et al. 2001). Thus, the CO emission lines can be used as a tracer of molecular hydrogen. The relationship between CO and $H_2$ can be stated as

$$N(\mathrm{H}_2) = X * I(\mathrm{CO}), \qquad (7)$$

where $N(\mathrm{H}_2)$ is the column density of $H_2$, $I(\mathrm{CO})$ is the surface brightness of CO emission, and $X$ is a conversion factor. The adopted method used to calculate the value of $X$ is that developed by Wilson (1995). The relationship between $X$ and the oxygen abundance is given by

$$\log \frac{X}{X_0} = 5.95 - 0.67[12 + \log(O/H)], \quad (8)$$

where $X_0 = 3 \times 10^{20}$ molecules cm$^{-2}$(K km s$^{-1}$)$^{-1}$. Since H$_2$ column densities derived in various published works are calculated using a variety of CO to H$_2$ calibrations, only original CO data were employed here. These data were then converted into H$_2$ column densities as a function of radius via Equations (7) and (8) using oxygen abundances derived from the gradients determined in Section 5.

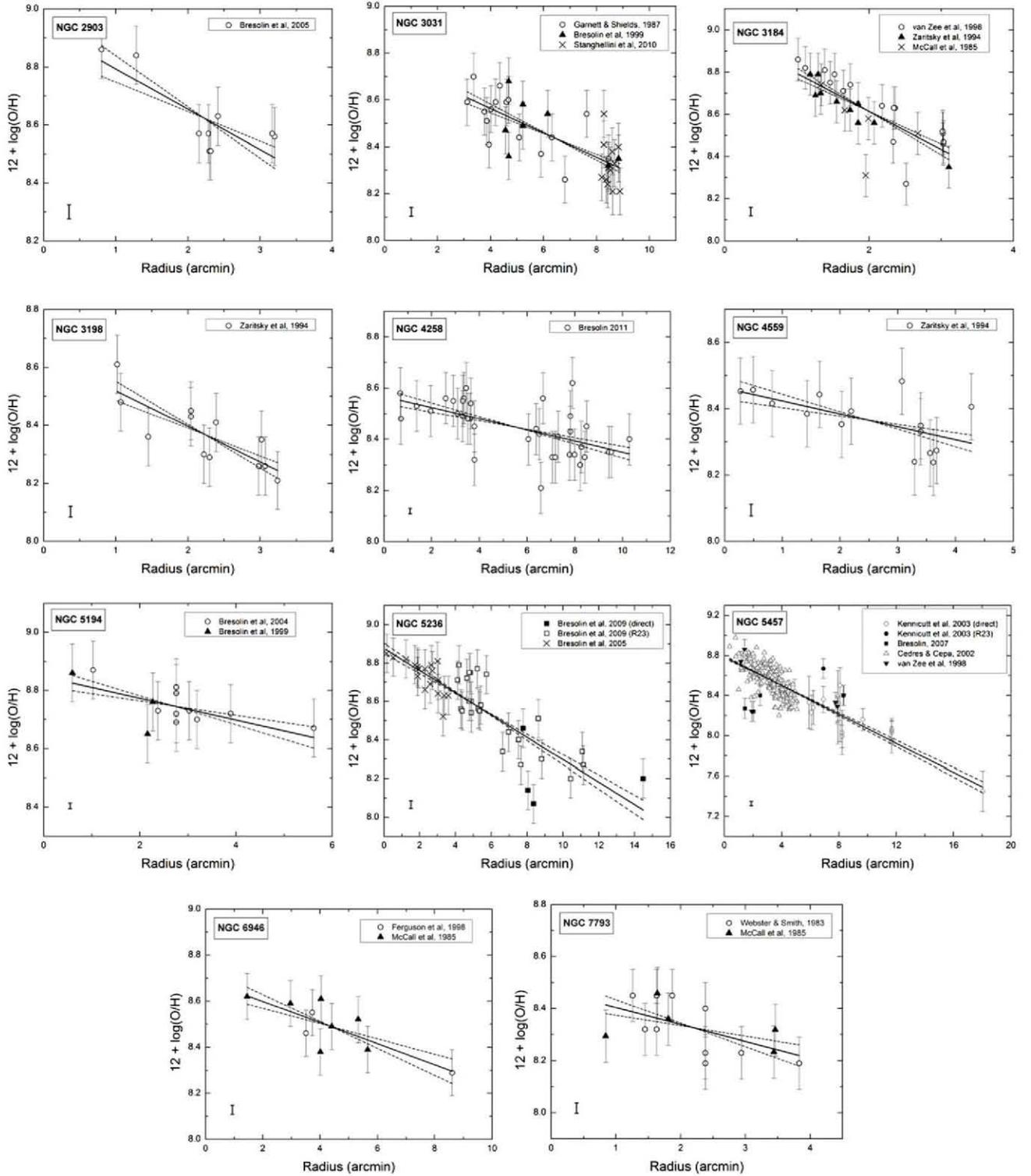

**Figure 3.** (Continued)

The column density of molecular hydrogen is known to decrease with radius at approximately an exponential rate (Young & Scoville 1982). Thus, graphs of the logarithm of the column density of H$_2$ molecules versus deprojected radius were plotted and a line was fit by least squares. This fit was used to define the approximate amount of H$_2$ at every radius.

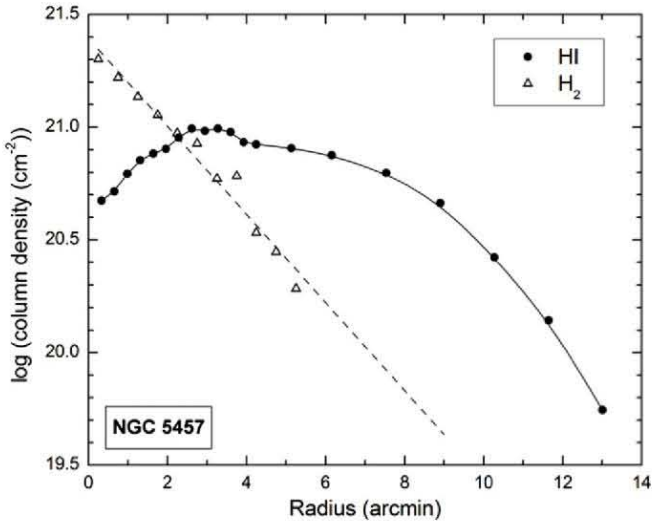

**Figure 4.** H I and H$_2$ column densities vs. deprojected radius across NGC 5457.

Data on molecular hydrogen were not available for all galaxies in this study. For those lacking coverage, the cumulative abundance is not truly "cumulative." However, as will be discussed in Section 7.3, the implications of not accounting for the contribution of H$_2$ are minimal.

Figure 4 shows typical radial trends of column densities of H I and H$_2$, specifically for NGC 5457. The column density scale is logarithmic, and the linear H$_2$ trend is displayed as a dashed line. Although the column density of H$_2$ starts at a higher level than that of H I, it quickly diminishes, leaving H I to dominate at large radii.

## 7. CUMULATIVE OXYGEN ABUNDANCES

### 7.1. Method of Calculation

The COA at a particular radius refers to the *total* number of oxygen atoms divided by the *total* number of hydrogen atoms within a disk of that radius. COAs were estimated as follows.

1. *Setting of annuli.* For each galaxy, deprojected radial intervals were chosen based on available H I data. "Rings" with a width as large as the H I radial intervals were generated. For example, in the case of NGC 5236, the disk was divided into 0.1 arcmin intervals (i.e., H I data were given for $r = 0.\!'1$, $0.\!'2$, $0.\!'3$, etc., where "$r$" is the deprojected radius). For this galaxy, a ring was made with boundaries at $r = 0.\!'5$ to $r = 1.\!'5$ and another was made with boundaries at $r = 1.\!'5$ to $r = 2.\!'5$, etc. For each of these rings, a total surface area was calculated.
2. $n(O)/n(H)$ *in each annulus.* At each radial interval (set by the H I column density data), the oxygen abundance was determined from the linear fit to the H II region measurements (see Figure 3).
3. *Q(H I) in each annulus.* The number of atoms of neutral hydrogen was calculated for each annulus by multiplying the H I column density at the corresponding radius by the surface area of the annulus.
4. *Q(H$_2$) in each annulus.* For those galaxies for which the molecular hydrogen contribution could be determined, the number of molecules of H$_2$ was found by multiplying the value of $N$(H$_2$) in each annulus interpolated from the gradient of log($N$(H$_2$)) versus radius by the surface area of the annulus.
5. *Q(H) in each annulus.* For those galaxies for which the H$_2$ contribution could not be determined (i.e., NGC 300, NGC 1232, NGC 1365, NGC 4559, and NGC 7793), the number of hydrogen atoms in each annulus was generated using only the atomic hydrogen density. For galaxies for which $N$(H$_2$) could be determined, the total number of hydrogen atoms was computed from $Q$(H I) + 2$Q$(H$_2$).
6. *Q(O) in each annulus.* Knowing the total hydrogen in each annulus, $Q$(H), as well as the oxygen abundance at every annulus, $n$(O)/$n$(H), calculating the number of atoms of oxygen abundance in every annulus was accomplished by multiplying the two together.
7. *Cumulative oxygen abundance out to radius r.* With knowledge of the $Q$(O) and $Q$(H) in every annulus, the COA out to any radius, $r$, was calculated by summing annular contributions to O and H within $r$.

Plots of the COA as a function of radius are shown in Figure 5 out to the radius of the last measurement of the H I column density. All cumulative abundances are expressed as 12+log (O/H). For easy comparison of trends, ordinates of all graphs span 0.5 dex in O/H (with the exception of NGC 3198). To show the extent of the extrapolation of the O/H gradient, vertical solid lines bound the range of O/H measurements. Galaxies for which the H$_2$ contribution could not be determined are notated accordingly.

The radius at which the curves start in Figure 5 is that at which the H I column density reaches a maximum. At smaller radii, the COA plots do not generally show any sort of overall trend. However, as more and more oxygen and hydrogen atoms are integrated, the trends become more stable and eventually tend to follow an exponentially decaying curve. In the H I plots of both NGC 598 and NGC 4258, the peak of the H I profile actually occurs at $r = 0$. In these cases, secondary peaks which occur at $r = 8.\!'5$ and $r = 7'$, respectively, have been used as starting points.

Two of the galaxies, NGC 1365 and NGC 5236, have prominent bars which seem to distort the COA trends at small radii. The presence of the bar within the inner part of each of these galaxies distorts the H I map and, in turn, oxygen abundance values. Therefore, for these two galaxies, the first points in the cumulative abundance plots have been chosen to be at radii at the ends of the bar.

### 7.2. The Asymptotic Value of O/H

The trends in the COA with radius in Figure 5 appear approximately exponential. This is especially evident for such galaxies as NGC 224, NGC 927, NGC 1232, NGC 2903, and NGC 4258, for which the curves reach an asymptote toward the last few points. However, not all galaxies trend to a clear asymptote. This inconsistency can be explained by the sampling of H I. The last point of each plot occurs at the radius of the last H I measurement available from the literature. It is suggested here that asymptotes would be clear for all galaxies if each one were sampled to large enough radii. Those galaxies which most closely approach an asymptote are used to define how to fit cumulative abundance curves universally.

It is the contention here that asymptotes of the plots in Figure 5 constitute good estimates of the overall oxygen abundances for most of the spiral galaxies in this study. Exponential functions were used to obtain asymptotes (and hence the overall

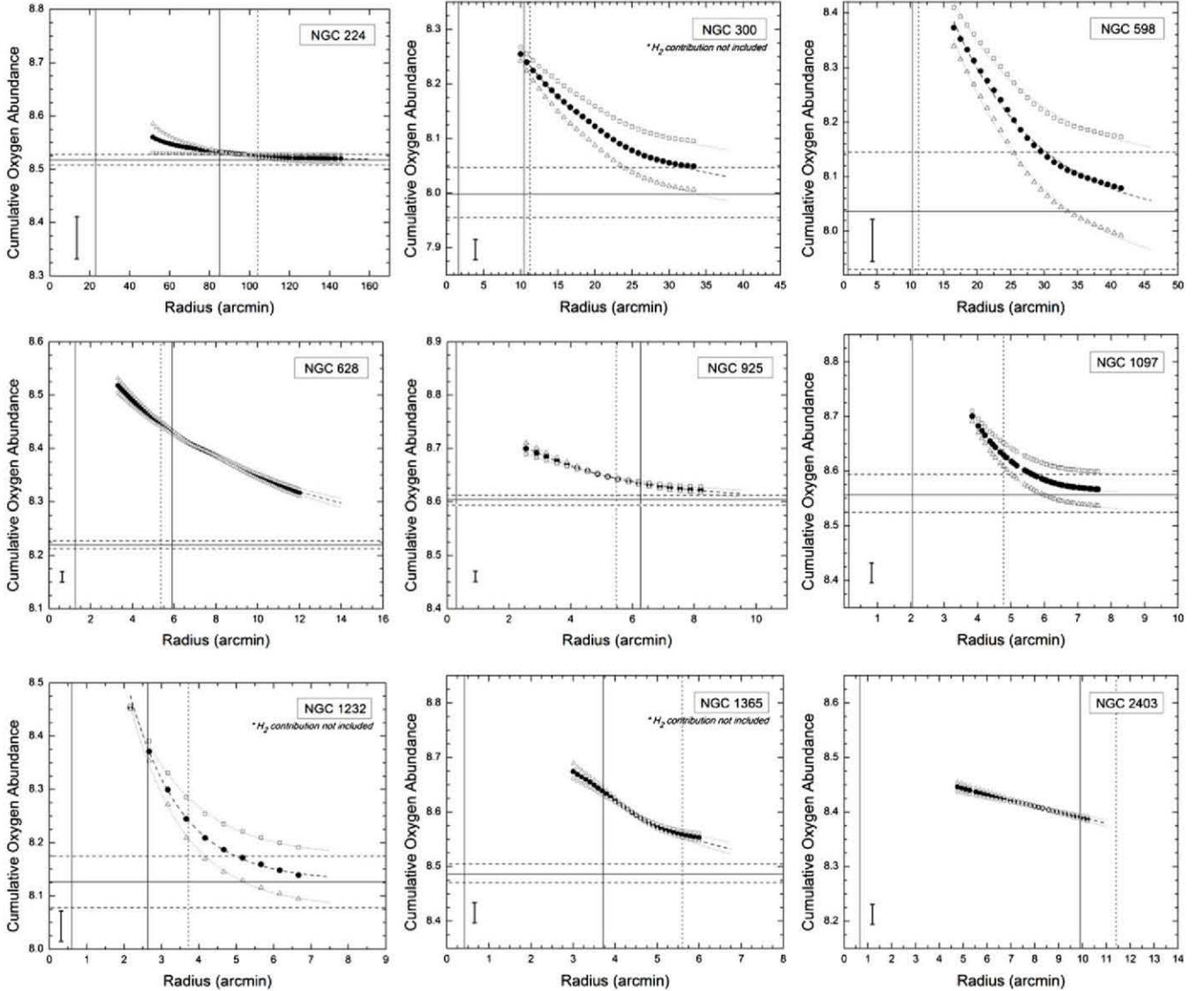

**Figure 5.** Cumulative oxygen abundance vs. deprojected radius for all galaxies in the sample. The dashed line passing through the closed circles represents an exponential fit. The solid horizontal line is the asymptote to the exponential fit, and it represents the overall oxygen abundance of the galaxy. The open triangles and open squares represent cumulative abundances using the maximum and minimum gradients allowed for the abundances (Figure 3), and the dashed horizontal lines represent their respective asymptotes. The vertical error in the fit arising from the uncertainties in the oxygen abundance level (Section 5) is shown in the bottom-left corner. The difference between the solid asymptote and the upper and lower bounds, added in quadrature to the uncertainty in the level, constitutes the error in the overall oxygen abundances. The data for NGC 2403 do not show any evidence of an approach to an asymptote, so only an upper limit to the overall abundance of this galaxy can be determined. For easy comparison, the range of all ordinate axes except for NGC 3198 has been fixed at 0.5 dex. The range of ordinates for NGC 3198 has been expanded to 0.7 dex in order to show all data points as well as the asymptote. The dotted vertical line represents the isophotal radius, $R_o$, as listed in Table 2. The two solid vertical lines bound the region in which oxygen abundances were measured.

abundances), since they generally fit well most of the COA trends with radius.

Beyond the peak in H I column density, the trend in the cumulative abundances with radius was modeled by

$$12 + \log(O/H) = Ae^{-r/\lambda} + b. \quad (9)$$

In Equation (9), $\lambda$ is the decay constant, $b$ is the asymptote, and $A + b$ is the nominal abundance at $r = 0$. As $r$ grows, and, hence, increasingly more of the total oxygen and hydrogen is included in the cumulative plot, the value of $12+\log(O/H)$ decays until it reaches the offset, $b$. Thus, the value of $b$ determines the oxygen abundance of the galaxy as a whole, i.e., the asymptotic oxygen abundance.

Table 4 provides a summary of the following abundance parameters for all galaxies in the sample: oxygen abundance gradient, asymptote of the cumulative abundances (corrected for missing $H_2$ as necessary), the COA at the last point sampled, and the difference in cumulative abundance between the last point sampled and the asymptote. The value of the asymptote is denoted here by $\Sigma(O)/\Sigma(H)$.

### 7.3. Contribution of Molecular Hydrogen

As shown in Table 2, data for molecular hydrogen were available for the galaxies NGC 224, NGC 598, NGC 628, NGC 925, NGC 1097, NGC 2403, NGC 2903, NGC 3031, NGC 3184, NGC 3198, NGC 4258, NGC 5194, NGC 5236, NGC 5457, and NGC 6946. Therefore, for these galaxies it was possible to compare COAs including molecular hydrogen to abundances that leave out the contribution from molecular hydrogen.

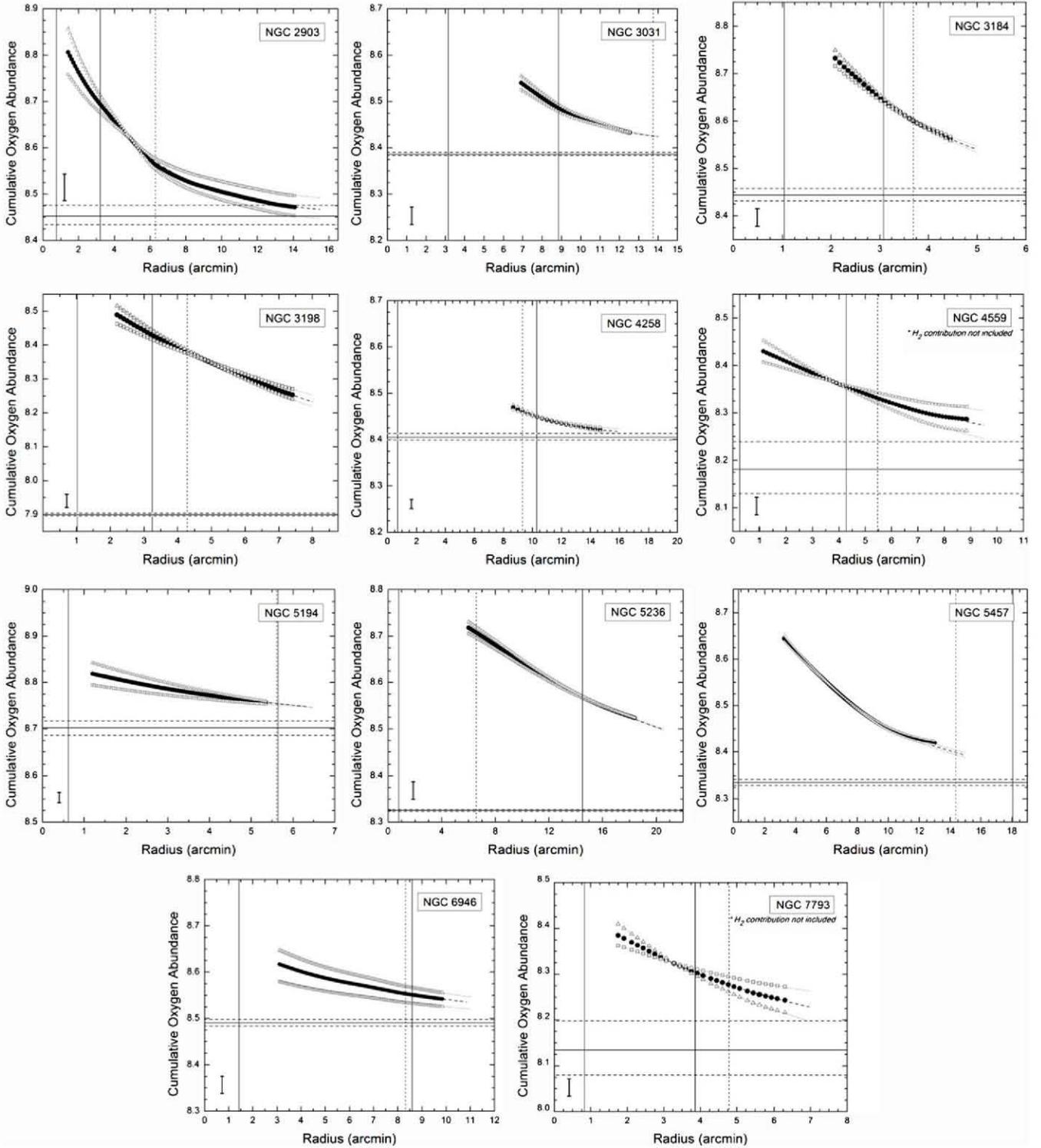

**Figure 5.** (Continued)

Figure 6 shows the COA versus radius for NGC 628 when molecular hydrogen is included (open circles) and when it is not included (solid squares). In this plot, the upper (solid) horizontal line represents the asymptote obtained when molecular hydrogen is included, while the lower (dashed) horizontal line represents the asymptote when it is not. The values for these two asymptotes are 8.22 and 8.13, respectively. Likewise, Sadavoy & McCall (2006) found differences for NGC 5457 and NGC 6946 to be 0.07 and 0.12, respectively.

Of the 15 galaxies for which $H_2$ data were available, only NGC 2903 showed a difference of greater than 0.09 dex between the asymptotic abundance with $H_2$ compared to the abundance not taking into account $H_2$. The difference between the asymptotes for NGC 2903 was 0.17 dex. No galaxy showed a decrease in the asymptotic oxygen abundance when molecular hydrogen was factored in. The average difference in asymptotes when $H_2$ is taken into account versus when it is not, when NGC 2903 is excluded, is 0.046 dex. When NGC 2903 is

Table 4
Parameters of the Oxygen Abundance Profiles

| Galaxy | O/H Gradient (dex kpc$^{-1}$) (1) | Asymptote ($\Sigma(O)/\Sigma(H)$) (2) | COA at Last Point (3) | COA (Last Point) $-\Sigma(O)/\Sigma(H)$ (4) | COA at $R_o$ (5) | COA ($R_o$) $-\Sigma(O)/\Sigma(H)$ (6) |
|---|---|---|---|---|---|---|
| NGC 224 | $-0.009 \pm 0.008$ | $8.52 \pm 0.04$ | $8.52 \pm 0.04$ | 0.00 | $8.53 \pm 0.01$ | 0.01 |
| NGC 300 | $-0.072 \pm 0.016$ | $8.05^a \pm 0.05$ | $8.10^a \pm 0.05$ | 0.05 | $8.23 \pm 0.02$ | 0.18 |
| NGC 598 | $-0.150 \pm 0.031$ | $8.04 \pm 0.11$ | $8.08 \pm 0.10$ | 0.04 | $8.18 \pm 0.09$ | 0.14 |
| NGC 628 | $-0.031 \pm 0.004$ | $8.22 \pm 0.02$ | $8.32 \pm 0.02$ | 0.10 | $8.45 \pm 0.01$ | 0.23 |
| NGC 925 | $-0.013 \pm 0.003$ | $8.60 \pm 0.02$ | $8.62 \pm 0.02$ | 0.02 | $8.64 \pm 0.01$ | 0.04 |
| NGC 1097 | $-0.027 \pm 0.006$ | $8.56 \pm 0.04$ | $8.57 \pm 0.04$ | 0.01 | $8.63 \pm 0.02$ | 0.07 |
| NGC 1232 | $-0.056 \pm 0.012$ | $8.15^a \pm 0.06$ | $8.19^a \pm 0.06$ | 0.01 | $8.25 \pm 0.04$ | 0.10 |
| NGC 1365 | $-0.022 \pm 0.004$ | $8.54^a \pm 0.03$ | $8.60^a \pm 0.03$ | 0.07 | $8.56 \pm 0.01$ | 0.02 |
| NGC 2403 | $-0.024 \pm 0.006$ | ... | $8.39 \pm 0.02$ | ... | ... | ... |
| NGC 2903 | $-0.057 \pm 0.015$ | $8.45 \pm 0.04$ | $8.47 \pm 0.04$ | 0.02 | $8.56 \pm 0.01$ | 0.11 |
| NGC 3031 | $-0.050 \pm 0.008$ | $8.39 \pm 0.02$ | $8.43 \pm 0.02$ | 0.05 | ... | ... |
| NGC 3184 | $-0.056 \pm 0.008$ | $8.44 \pm 0.02$ | $8.56 \pm 0.02$ | 0.12 | $8.60 \pm 0.01$ | 0.16 |
| NGC 3198 | $-0.031 \pm 0.007$ | $7.90 \pm 0.02$ | $8.25 \pm 0.03$ | 0.35 | $8.38 \pm 0.01$ | 0.48 |
| NGC 4258 | $-0.010 \pm 0.002$ | $8.41 \pm 0.02$ | $8.42 \pm 0.02$ | 0.02 | $8.46 \pm 0.01$ | 0.05 |
| NGC 4559 | $-0.016 \pm 0.006$ | $8.23^a \pm 0.06$ | $8.34^a \pm 0.03$ | 0.11 | $8.33 \pm 0.01$ | 0.10 |
| NGC 5194 | $-0.017 \pm 0.006$ | $8.70 \pm 0.02$ | $8.76 \pm 0.02$ | 0.05 | ... | ... |
| NGC 5236 | $-0.043 \pm 0.004$ | $8.33 \pm 0.02$ | $8.52 \pm 0.02$ | 0.20 | $8.71 \pm 0.01$ | 0.38 |
| NGC 5457 | $-0.035 \pm 0.002$ | $8.34 \pm 0.01$ | $8.42 \pm 0.01$ | 0.08 | ... | ... |
| NGC 6946 | $-0.027 \pm 0.007$ | $8.49 \pm 0.02$ | $8.54 \pm 0.03$ | 0.05 | $8.55 \pm 0.02$ | 0.06 |
| NGC 7793 | $-0.086 \pm 0.024$ | $8.18^a \pm 0.07$ | $8.29^a \pm 0.04$ | 0.11 | $8.28 \pm 0.02$ | 0.10 |

**Notes.** (1) Oxygen abundance gradient (see Figure 3). (2) Asymptote of the cumulative abundance profile (see Figure 5). (3) Cumulative oxygen abundance at the last measurement in the cumulative abundance profile. (4) Difference between cumulative oxygen abundance at last point and asymptote. (5) Cumulative oxygen abundance at the isophotal radius, $R_o$. (6) Difference between cumulative oxygen abundance at $R_o$ and asymptote.
$^a$ 0.05 dex has been added to account for the systematic effect of missing $H_2$ data.

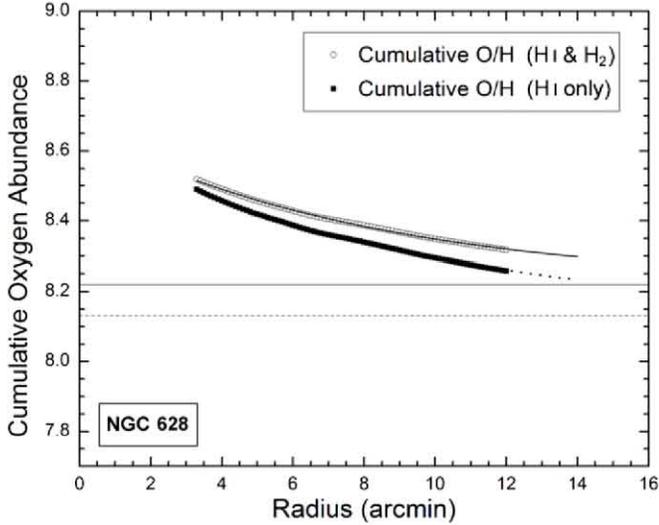

**Figure 6.** Cumulative oxygen abundance vs. deprojected radius for NGC 628, with and without the contribution of $H_2$. The upper points, exponential fit, and solid horizontal line are equivalent to those in Figure 5, i.e., they apply to the cumulative abundances including $H_2$. The lower points are the cumulative abundances of oxygen including atomic hydrogen atoms only. The dotted line running through the curve and the dashed horizontal line represent the exponential fit and the asymptote, respectively.

included, the mean difference is 0.055 dex. Since this is a systematic effect, final cumulative abundances for all galaxies lacking $H_2$ data were determined by adding 0.05 dex to the asymptote of the cumulative abundances.

### 7.4. Uncertainties in Asymptotes

The robustness of this method for evaluating asymptotic abundances was tested by comparing the result for an exponentially decaying curve beginning at a radius 10% greater than that of the peak of the H I column density with that for a curve starting at the H I peak. For all galaxies in the sample, the maximum difference was less than 0.01 dex. This difference is less than the error associated with the each asymptote.

Pilyugin (2002) showed that for spiral galaxies, no bend in the slope of the oxygen abundance gradient exists. In fact, it was assumed here that the gradient is constant across the entire extent of H I sampled. For all galaxies in the sample, a comparison was made between asymptotes that were calculated with gradients whose slopes stay constant, and those that were calculated with gradients that flatten out at 12+log(O/H) = 7.2. The value of 7.2 is the oxygen abundance of I Zw 18, one of two of the most metal-deficient star-forming galaxies known (Thuan & Izotov 2005). For all galaxies in the sample, the difference in asymptotes is less than 0.01 dex. This is much smaller than the error contributed already by the uncertainty in the fit to the oxygen abundance gradient.

Some of the cumulative abundance trends demonstrate slight deviations from exponentials that can lead to the suspicion that the asymptotes obtained are systematically too low. For example, for NGC 300, NGC 4559, and NGC 5457, it can be seen that the cumulative abundances at the largest radii begin to level out faster than the exponential does. For these galaxies, the asymptote of the exponential function may be a lower limit to the true overall cumulative abundance. An upper limit is given by the point at the largest radius. In 17 of the 20 galaxies, the asymptote and the last point are within 0.11 dex of each other. Notable exceptions are NGC 2403, NGC 3198, and NGC 5263.

Several measures could be taken to improve the confidence of the asymptotes. Increasing the range of radii over which oxygen abundances are measured would reduce extrapolation uncertainties. Also, H I maps that extend to larger radii would allow for an extension to the cumulative abundance plots. Since many of the curves only begin to level out toward the edges of the H I maps, sampling H I out farther will likely lead to clearer asymptotic behavior for the cumulative abundances. In this work, the exponential fit to the cumulative abundance plots starts at the radius at which the column density of H I peaks. For most of the galaxies, this choice appears to be valid. However, for others, it appears that another starting point may in fact lead to a better fitting exponential.

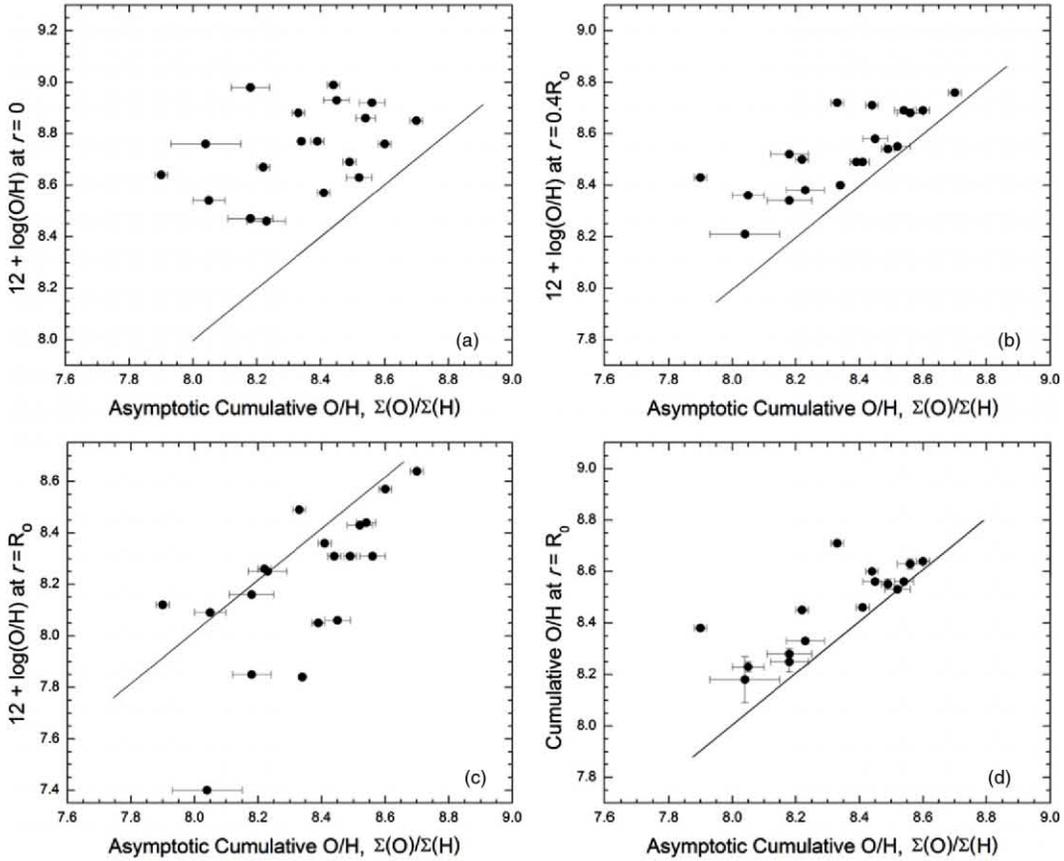

**Figure 7.** Comparison of abundances at fiducial radii with asymptotic cumulative abundances. Solid lines mark equality. (a) O/H at $r = 0$; (b) O/H at $r = 0.4\ R_o$; (c) O/H at $r = R_o$; (d) Cumulative O/H at $r = R_o$.

### 7.5. Cumulative Abundance of NGC 2403

Although the radial trend of COAs leads to an asymptotic abundance for most spiral galaxies, this is not the case for NGC 2403. An exponential function cannot be said to approximate the shape of the COA plot for this galaxy. The reason why the method fails for NGC 2403 is probably due to the fact that it possesses a very shallow abundance gradient. When integrating out to larger and larger radii, the cumulative abundance decreases very slowly with radius. This same shortcoming is observed (though to a much lesser extent) for some of the other shallow-gradient galaxies, such as NGC 925, NGC 4258, and NGC 5194.

For NGC 2403, a better approximation to the overall abundance may be the cumulative abundance at the limiting radius of measurement. However, this value has to be considered to be an upper limit.

### 7.6. Oxygen Abundance at $r = 0$, $r = 0.4\ R_o$, and $r = R_o$

Zaritsky et al. (1994) proposed that overall abundances for spirals be quantified by O/H at $0.4\ R_o$, where $R_o$ is the isophotal radius. Values of $R_o$ are listed in Table 2. Table 5 and Figure 7(b) compare abundances at $r = 0.4\ R_o$ and the asymptotic cumulative abundances, $\Sigma(O)/\Sigma(H)$, for all galaxies in the sample. The diagonal line in Figure 7(b) represents equality. The correlation is weak. In all instances, the abundance at $0.4\ R_o$ is larger than the asymptotic abundance. The average difference between the two is 0.22 dex. Clearly, the abundance at $0.4\ R_o$ is an unreliable index of the overall oxygen abundance of a spiral.

The Zaritsky method for calculating overall oxygen abundances suffers from several inadequacies. The method necessarily relies on an estimation of metallicity for the whole galaxy based on the oxygen abundance at a radius defined by a particular surface brightness. The choice of surface brightness is based on a comparison of the oxygen abundance at $0.4\ R_o$ versus blue magnitude, circular velocity, and Hubble type. However, no proof has been found to directly relate the overall oxygen abundance to any of these parameters. Moreover, no standard shape for H I column density distributions has been observed. Therefore, absolute amounts of both hydrogen and oxygen at radii larger than $0.4\ R_o$ can vary unpredictably, leading to a COA uncoupled from O/H at $0.4\ R_o$.

Table 5 and Figure 7 also compare oxygen abundances at the fiducial radii $r = 0$ and $r = R_o$ with the asymptotic cumulative abundances. Again, correlations are weak. It is concluded that it is not possible to reliably gauge the global chemical composition of a spiral disk from a measurement of the oxygen abundance at a single radius.

Table 4 and Figure 7(d) compare the cumulative abundances at $r = R_o$ with the asymptotic cumulative abundances. Figure 7(d) shows a tighter correlation than that of $\Sigma(O)/\Sigma(H)$ and the oxygen abundances at either $r = 0$, $r = 0.4\ R_o$,

**Table 5**
Comparison of Abundances at Fiducial Radii with Asymptotic Cumulative Abundances

| Galaxy | $0.4\, R_O$ (arcmin) | O/H at $r = 0$ | O/H at $r = 0.4\, R_O$ | O/H at $r = R_O$ | Asymptote ($\Sigma(O)/\Sigma(H)$) | O/H at $0.4\, R_O$ $-\Sigma(O)/\Sigma(H)$ |
|---|---|---|---|---|---|---|
| NGC 224  | 38.11 | 8.63 | 8.55 | 8.43 | $8.52 \pm 0.04$ | 0.04 |
| NGC 300  | 4.38  | 8.54 | 8.36 | 8.09 | $8.05^a \pm 0.05$ | 0.32 |
| NGC 598  | 3.72  | 8.76 | 8.21 | 7.40 | $8.04 \pm 0.11$ | 0.58 |
| NGC 628  | 1.96  | 8.67 | 8.50 | 8.26 | $8.22 \pm 0.02$ | 0.30 |
| NGC 925  | 2.10  | 8.76 | 8.69 | 8.57 | $8.60 \pm 0.02$ | 0.08 |
| NGC 1097 | 1.87  | 8.92 | 8.68 | 8.31 | $8.56 \pm 0.04$ | 0.13 |
| NGC 1232 | 1.48  | 8.98 | 8.52 | 7.85 | $8.18^a \pm 0.06$ | 0.35 |
| NGC 1365 | 2.24  | 8.86 | 8.69 | 8.44 | $8.54^a \pm 0.03$ | 0.11 |
| NGC 2403 | 4.38  | 8.50 | 8.41 | 8.27 | ... | ... |
| NGC 2903 | 2.52  | 8.93 | 8.58 | 8.06 | $8.45 \pm 0.04$ | 0.12 |
| NGC 3031 | 5.38  | 8.77 | 8.49 | 8.05 | $8.39 \pm 0.02$ | 0.10 |
| NGC 3184 | 1.48  | 8.99 | 8.71 | 8.31 | $8.44 \pm 0.02$ | 0.27 |
| NGC 3198 | 1.70  | 8.64 | 8.43 | 8.12 | $7.90 \pm 0.02$ | 0.53 |
| NGC 4258 | 3.72  | 8.57 | 8.49 | 8.36 | $8.41 \pm 0.02$ | 0.08 |
| NGC 4559 | 2.14  | 8.46 | 8.38 | 8.25 | $8.23^a \pm 0.06$ | 0.15 |
| NGC 5194 | 2.16  | 8.85 | 8.76 | 8.64 | $8.70 \pm 0.02$ | 0.06 |
| NGC 5236 | 2.58  | 8.88 | 8.72 | 8.49 | $8.33 \pm 0.02$ | 0.40 |
| NGC 5457 | 5.77  | 8.77 | 8.40 | 7.84 | $8.34 \pm 0.01$ | 0.02 |
| NGC 6946 | 2.30  | 8.69 | 8.54 | 8.31 | $8.49 \pm 0.02$ | 0.09 |
| NGC 7793 | 1.87  | 8.47 | 8.34 | 8.16 | $8.18^a \pm 0.07$ | 0.16 |

**Note.** [a] 0.05 dex has been added to account for the systematic effect of missing $H_2$ data.

or $r = R_o$. The COAs at $r = R_o$ are systematically higher than $\Sigma(O)/\Sigma(H)$ by $0.14 \pm 0.12$ dex. If the two outliers (NGC 3198 and NGC 5236) are removed, the difference becomes $0.10 \pm 0.06$ dex.

The new method discussed in this study takes into account absolute amounts of both oxygen and hydrogen at all radii. Therefore, it calculates the true ratio of oxygen atoms to hydrogen atoms. It is not based on an estimation of oxygen abundance at any specific radius. Thus, it is justified to conclude that the method described in this study gives a more accurate estimation of the overall oxygen abundance of a spiral.

### 8. CONCLUSIONS AND FUTURE DIRECTIONS

In this study, a new method for calculating the total oxygen abundance of spiral galaxies, first introduced by Sadavoy & McCall (2006), has been established and improved through application to 20 spirals. This method involves finding the cumulative amounts of oxygen and hydrogen at increasing radii. The COA approaches an asymptote at large radii. At radii beyond the peak in the column density of H I, the trend in COAs can be approximated by an exponential function, and the asymptote can be used to gauge the overall oxygen abundance of the galaxy.

The bulk of the hydrogen gas within each galaxy is contained in two forms: neutral hydrogen, H, and molecular hydrogen, $H_2$. From analysis of 15 galaxies, it was found that molecular hydrogen only contributes a minimal amount to the overall hydrogen content when variations in $N(H_2)/I(CO)$ with O/H are accommodated. Consequently, of the 15 galaxies with CO data available, the average difference in COA from when $H_2$ was taken into account compared to when it was not was only 0.05 dex. Thus, COAs for galaxies for which $H_2$ data are not available should be augmented by 0.05 dex to account for the missing hydrogen.

The method developed in this thesis is the best technique for finding overall oxygen abundances of galaxies displaying radial gradients in chemical composition, particularly spirals. It is unambiguous in the way it estimates abundances, since it accounts for all atoms of oxygen and hydrogen without linkage to an arbitrary choice of radius. Other techniques for estimating the overall oxygen abundance, most notably the estimate of $0.4\, R_o$ proposed by Zaritsky et al. (1994), are unreliable because they are explicitly tied to radius. While the estimate of $0.4\, R_o$ may sometimes give a good approximation, it has been shown that a difference between the $0.4\, R_o$ method and the one presented here can be more than 0.5 dex (see Table 5). Each galaxy is unique in its distribution of oxygen and hydrogen, and there is no clear evidence to show that the overall oxygen abundance can be gauged from a measurement at a single radius.

Despite its success, the new method for calculating oxygen abundances is not without limitations. In this study, NGC 2403 did not demonstrate an exponentially decaying profile when cumulative abundances were calculated out to large radii. This could be explained by its very shallow oxygen abundance gradient, similar to dwarf irregular galaxies. This shallow gradient contributed to an almost flat cumulative abundance curve. It might even be reasonable to approximate the oxygen abundance of this galaxy to be constant across all radii. In this case, the cumulative abundance curve would also be flat, and the global abundance would be obvious.

The method developed in this study will be used in a subsequent paper in an effort to compare the chemical evolution of spiral galaxies to that of dwarf irregulars.

The authors thank Raoul Haschke for his early contributions to the determination of the cumulative abundance of M33. M.L.M. is grateful to the Natural Sciences and Engineering Research Council of Canada for its continuing support.